\documentclass[reprint,superscriptaddress,amsmath,amsfonts,amssymb,floatfix]{revtex4-2}

\usepackage{graphicx,xcolor,latexsym}
\usepackage[T1]{fontenc}
\usepackage[utf8]{inputenc}
\usepackage[final]{microtype}

\newcommand{\MyPaperTitle}{Dynamics of a Field Emitted Beam from a Microscopic Inhomogeneous Cathode}

\usepackage[]{url}
\usepackage[colorlinks=true, linkcolor=blue, citecolor=blue, urlcolor=blue, bookmarks=true, breaklinks=true,
            pdftitle={\MyPaperTitle},
            pdfauthor={Kristinn Torfason}]{hyperref}
\usepackage{bookmark}
\usepackage[]{subcaption}

\newcommand{\ud}{\mathrm{d}}

\usepackage[normalem]{ulem}

\usepackage{tikz}
\usetikzlibrary{patterns,decorations.pathmorphing,decorations.markings,shapes,arrows,positioning,decorations.pathreplacing,calc,backgrounds,external,spy}

\def\clap#1{\hbox to 0pt{\hss#1\hss}}

\begin{document}
 

\title{\MyPaperTitle}


\author{Kristinn Torfason}
\homepage[]{http://nano.ru.is}
\affiliation{Department of Engineering, Reykjavik University, Menntavegur 1, IS-102 Reykjavik, Iceland}
\author{Anna Sitek}
\affiliation{Department of Engineering, Reykjavik University, Menntavegur 1, IS-102 Reykjavik, Iceland}
\affiliation{Department of Theoretical Physics, Wroclaw University of Science and Technology, 50-370 Wroclaw, Poland}
\author{Andrei Manolescu}
\affiliation{Department of Engineering, Reykjavik University, Menntavegur 1, IS-102 Reykjavik, Iceland}
\author{\'Ag\'ust Valfells}
\affiliation{Department of Engineering, Reykjavik University, Menntavegur 1, IS-102 Reykjavik, Iceland}

\date{\today}

\begin{abstract}
We investigate by molecular dynamics simulations~\cite{doi:10.1063/1.4914855, doi:10.1063/1.4972821} a beam of electrons released via field emission from a planar cathode surface of 1 ${\rm \mu m}^2$  with an inhomogeneous two-level work function, $\phi_{\rm low}$ and $\phi_{\rm high}$. A rectangular grid, where each cell can have one out of two values of the work function, is used as a model. The number of cells in the grid ranges from 6x6 to 96x96. We compare a periodic checkerboard arrangement with disordered distributions of patches. We perform multiple simulations and randomize the pattern each time. We study the beam behavior by calculating the position and velocity of each electron, r.m.s. emittance, and the brightness of the electron beam.
The emittance increases while brightness decreases,
with the mean distance between patches with $\phi_{\rm low}$ when they are in minority, and they switch the behavior vs. the mean distance between patches with $\phi_{\rm high}$ when these patches are in minority, respectively.
The Coulomb interaction between all particles is fully included in our simulations.
\end{abstract}

\maketitle

\section{Introduction\label{sec:intro}}
Field emitted electrons have a wide range of use, such as in microwave tubes, electron microscopes,
vacuum channel transistors~\cite{stoner2012nanoelectronics, doi:10.1063/1.4717751}, ultrafast diodes~\cite{zhang2016ultrafast, RevModPhys.92.025003, FEIST201763}, and space propulsion systems~\cite{2000mfss.conf..271M}. For many of these applications, it is important to understand and control the electron beam dynamics. The dynamics of the beam are strongly affected by the conditions at the point of emission, and hence the emitting surface influences the properties of the electron beam greatly. The effects of the cathode surface on the beam can be via the shape of the surface~\cite{PhysRevSTAB.17.043402,doi:10.1063/1.4972821}, roughness or imperfections~\cite{doi:10.1063/1.5097149, doi:10.1063/1.338833, krasilnikov2006impact}, temperature~\cite{Ilkov2015}, or material properties such as the work function~\cite{doi:10.1116/1.2827508}.

In practice, a cathode is never completely homogeneous and includes a certain amount of randomness or disorder.  It is therefore important to evaluate the effects of an imperfect cathode on the quality of the beam. In particular low emittance and high brightness are often desired in many electron devices~\cite{krasilnikov2006impact, OShea1853, 989967}. Jensen et~al.~\cite{doi:10.1116/1.2827508} used a general emission model, incorporating thermal, field, and photoemission, and a particle-in-cell approach to study the consequences of cathode nonuniformity due to work function variation. One of the findings was that the beam emittance is more sensitive to such imperfections than the current itself. In a recent paper, we considered a microscopic vacuum diode with a field emission mechanism and an inhomogeneous cathode comprised of four active patches separated by inactive areas~\cite{Haraldsson2020}. We found that the mutual space-charge effects set in rapidly when the distance between the emitting patches is considerably shorter than the diode gap. 

In the present work, we consider a microscopic planar diode, with a square emitting area on the cathode split into checks with different work functions. We study how the work function distribution affects the current, the emittance, and the brightness.
We build upon previous work~\cite{doi:10.1063/1.4914855, doi:10.1063/1.4972821}, using our Molecular Dynamics (MD) code where the Coulomb interaction between all electrons is fully incorporated. The MD approach takes the discrete emission of electrons into account and allows us to track individual electron trajectories. This allows us to calculate quantities that characterize the beam quality, such as velocity, density, current, emittance, and brightness. The simulations use both designed surfaces and randomly generated ones for the cathode.

\autoref{sec:method} gives a description of the methodology and the model used. Results are
presented in \autoref{sec:results} with a short summary in \autoref{sec:summary}.
\section{Methodology\label{sec:method}}
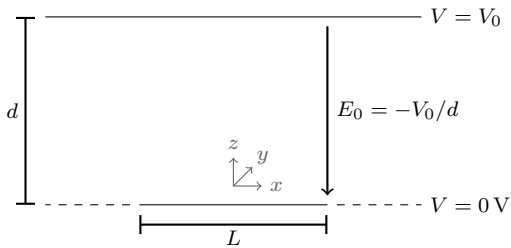
\begin{figure}[bt]
  \centering
  %
\begin{tikzpicture}[font=\footnotesize, scale=0.50]
  \draw[] (-5.0, 5.0) -- (5.0, 5.0) node[right] {\(V = V_0\)};
  
  \draw[] (-2.5, 0.0) -- (2.5, 0.0);
  \draw[dashed] (2.75, 0.0) -- (5.0, 0.0) node[right] {\(V = 0\,\mathrm{V}\)};
  \draw[dashed] (-2.75, 0.0) -- (-5.0, 0.0);
  \draw[|-|, thick, yshift=-15pt] (2.5, -0.0) -- (-2.5, -0.0) node [black, midway, yshift=-5pt] {\(L\)};
  
  \draw[|-|, thick, xshift=-15pt, yshift=0pt] (-5.0, 0.0) -- (-5.0, 5.0) node [black, midway, xshift=-5pt] {\(d\)};
  
  \draw[->, thick] (2.5, 4.75) -- (2.5, 0.25) node[midway, right] {\(E_{0} = - V_0/d\)};
  
  \draw[black!62.5!white, ->] (0.0, 0.5) -- (0.0, 1.25) node[above] {\(z\)};
  \draw[black!62.5!white, ->] (0.0, 0.5) -- (0.75, 0.5) node[right] {\(x\)};
  \draw[black!62.5!white, ->] (0.0, 0.5) -- (0.5, 1.0) node[right, xshift=-1.5pt, yshift=3.5pt] {\(y\)};
\end{tikzpicture}
  \caption{The model of the planar nanodiode.}
  \label{fig:system}
\end{figure}

\autoref{fig:system} shows the system being modeled, a planar vacuum diode with gap
spacing \(d\). The cathode in the system is grounded, and the potential \(V_0\)
is applied to the anode, which creates an electric field that is strong enough to cause field emission.
The emission is only allowed to take place on a square area with side lengths \(L\) and is described by the Fowler-Nordheim law. The equation for it is
\begin{equation}
  J = \frac{A}{t^2(\ell)\phi}F^2 \mathrm{e}^{-\nu(\ell)B\phi^{\frac{3}{2}}/F}\, ,
\end{equation}
where \(\phi\) is the work-function and \(F\) is the field at the surface of the cathode, taken to be positive. \(A = e^2/(16\pi^2\hbar)\; [\mathrm{A}\,\mathrm{eV}\,\mathrm{V}^{-2}]\)
and \(B = 4/(3\hbar) \sqrt{2m_e e}\; [\mathrm{eV}^{-\frac{3}{2}}\,\mathrm{V}\,\mathrm{m}^{-1}]\)
are the first and second Fowler-Nordheim constants, while \(\nu(\ell)\) and \(t(\ell)\)
are called the Nordheim functions and arise due to the image-charge effect~\cite{Forbes08112007}.

In this work, the emission area on the cathode has a work function that varies with position, \(\phi(x,\,y)\). A rectangular grid is created where each
square check can have its own work function.
A Metropolis-Hastings (MH) like algorithm is used to determine where
electron emission occurs. The exponential part
\begin{equation}\label{eq:prob}
    D_F = \exp \left(-\nu(\ell)B\phi^{\frac{3}{2}}/F \right)\, ,
\end{equation}
of the Fowler-Nordheim function is interpreted as an escape probability for the electrons, and is used as a
probability distribution for the MH algorithm. This makes both the field and the work function
affect the placement of electrons on the cathode.
The part in front of the exponent
\begin{equation}
    J_{\rm supp} = \frac{A}{t^2(\ell)\theta}F^2\, ,
\end{equation}
is interpreted as the electron supply. It is integrated~\cite{HAHN200578} over the surface of the cathode to obtain the number of electrons coming to the surface.
Once a suitable location for emission has been found for en electron using the MH algorithm, a uniform random number is generated between 0 and 1. That number is then compared with \autoref{eq:prob} to determine if emission occurs at that location or not.

We use the Velocity-Verlet algorithm to update each electron's position and velocity in every time step of the simulation. The force on each electron in the diode gap is calculated 
using Coulomb's law. From these three quantities, the code can calculate various properties of the electron beam. Such as current, emittance, brightness, and density.
The current is calculated using the Shockley–Ramo theorem~\cite{doi:10.1063/1.1710367, 1686997}. Where we sum over the contribution from all electrons to the total current. The equation is
\begin{equation}
    I = \frac{q}{d}\sum_i v_{z,\, i}\, ,
\end{equation}
where \(q\) is the electron charge, \(v_z\) the \(z\)-component of the instantaneous velocity (with \(z\) being normal to the cathode), and \(d\) the gap spacing.
To describe the spread of the electron beam at the anode, we calculate the transverse emittance of the beam in the \(x\)- and \(y\)-directions.
We use the statistical emittance~\cite{reiser1994theory, buon1992beam} calculated using,
\begin{equation}\label{eq:emittance}
  \epsilon_x = \sqrt{\left< x^2 \right> \left< x^{\prime 2} \right> - \left< xx^\prime \right>^2}\, .
\end{equation}
The \(x\)-position and the slope of the electron trajectory \(x^\prime = \ud x / \ud z \approx v_x / v\) are recorded when the electrons pass through the anode.
The slope is calculated using the velocity of the electrons. The emittance in the \(y\)-direction is calculated in the same way.
Another quality factor characterizing the beam is brightness~\cite{reiser1994theory} which
is calculated from the emittance and current using
\begin{equation}\label{eq:brightness}
    B = \frac{2I}{\pi^2\epsilon_x\epsilon_y}\, .
\end{equation}
Where \(I\) is the current and \(\epsilon_{x,\,y}\) are the emittance in the \(x\)- and \(y\)-directions, respectively.

\section{Results\label{sec:results}}
All the simulations were performed for the gap voltage and gap spacing fixed at \(V_0 = 2\,\mathrm{kV}\) and  \(d = 1\,\mathrm{\mu m}\), respectively. The emitter side length is set to \(L = 1\,\mathrm{\mu m}\). We studied systems with two different work function values, which we set to $\phi_{\rm low}=2.0$~eV and $\phi_{\rm high}=2.5$~eV,
forming either a checkerboard pattern on the cathode's surface or randomly distributed patches on a square lattice.

\subsection{Checkerboard pattern}
  \begin{figure}
    \centering
    {\phantomsubcaption\label{fig:checkerboard_emittance-a}}
    {\phantomsubcaption\label{fig:checkerboard_emittance-b}}
    \includegraphics[]{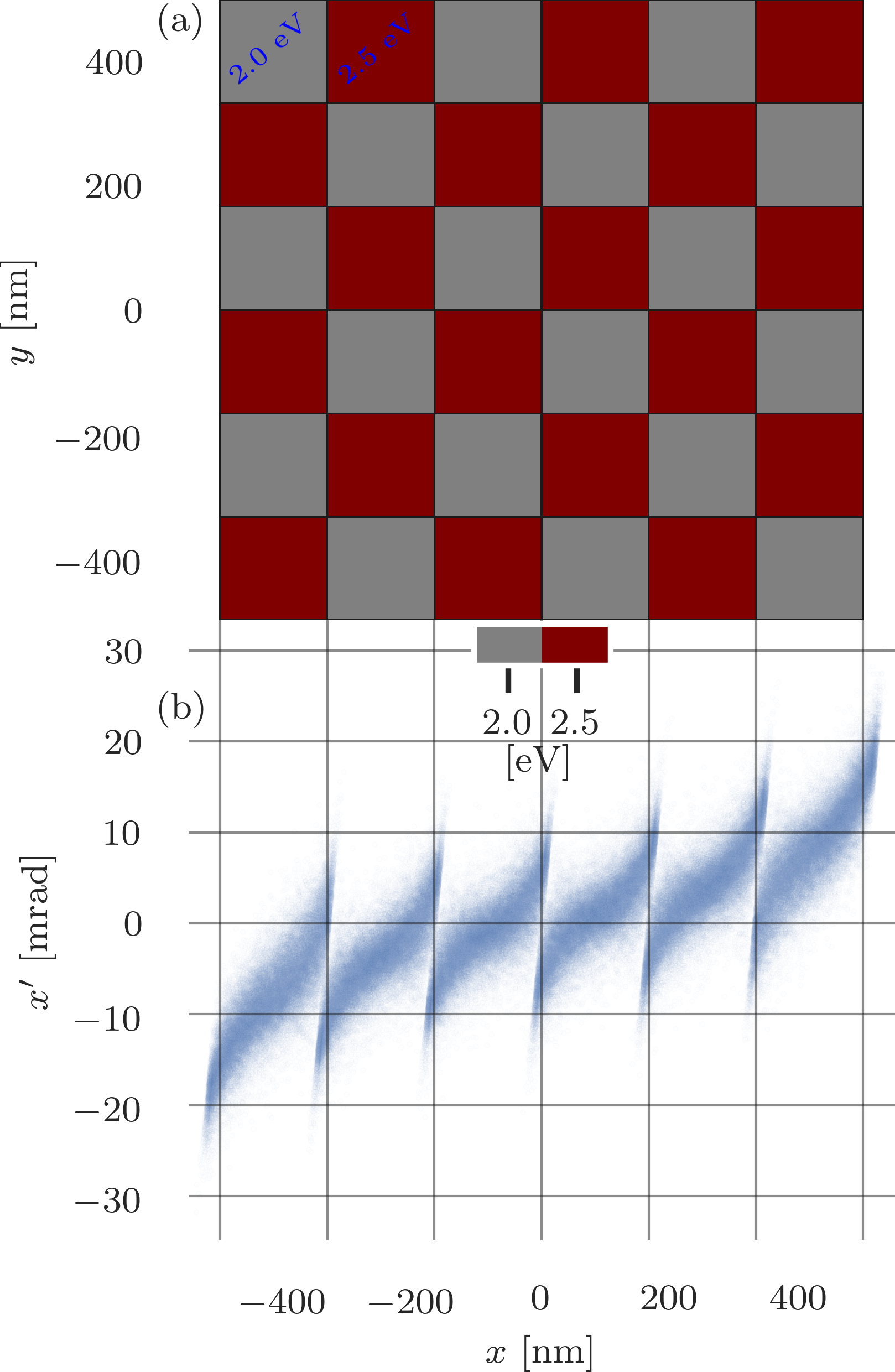}
    \caption{(a) The 6x6 checkerboard pattern used in the simulations. \textcolor{red}{Red} squares indicate the areas with the work function value 2.5~eV and \textcolor{gray}{gray} ones with 2.0~eV. (b) The \(x\)-\(x^\prime\) coordinate phase space diagram of the electron beam.
    }
    \label{fig:checkerboard_emittance}
  \end{figure}
\autoref{fig:checkerboard_emittance} shows the emitting area of the cathode with a \(6\times 6\) checkerboard pattern 
with the corresponding phase space diagram for \(x\)\nobreakdash-\(x^\prime\), shown below. As seen in \autoref{fig:checkerboard_emittance-b} each column forms its own S-shaped pattern in the phase space, i.e., locally every column acts as a single periodic emitter with its own emittance that contributes to the total emittance. The angular range of the S-shaped pattern depends on the column's position with respect to the center of the cathode. The electrons originating from the central columns are symmetric in phase space with respect to line \(x^\prime =0\) and span the shortest range of angles (slightly exceeding 20 mrad). The angular dispersion of the S-shaped pattern increases with the distance from the cathode's center. It is shifted towards higher or lower values depending on whether the column is situated to the right or left of the cathode center, respectively. The electrons from the edge columns are unconstrained by the vacuum outside the boundary of the emitter, while the spread in the opposite direction is hindered by the particles emitted from the internal columns. As a result, the angular spread outward from the edge is much larger than inward from the edge.
Towards the center of the cathode the space-charge becomes more symmetric, and thus the S-shaped patterns become more symmetric. The corresponding phase-space diagram for the y coordinate (\(y\)\nobreakdash-\(y^\prime\)) looks the same and, therefore, is not shown.
\begin{figure}
  \centering
  \includegraphics[]{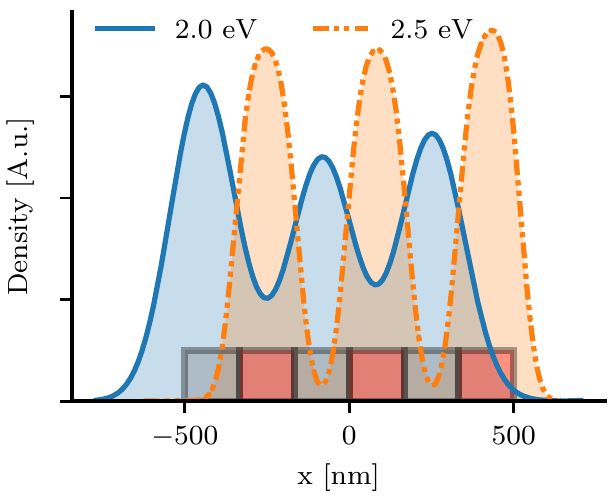}
  \caption{Electron density through a single row on the checkerboard in \protect\autoref{fig:checkerboard_emittance}.
  The density for 2.0~eV (\protect\textcolor{blue}{blue} solid line)
  and 2.5~eV (\protect\textcolor{orange}{orange} dashed line)
  are both scaled such that the area under each curve is 1.
  The boxes at the bottom of the graph represent the checkerboard row.
  \protect\textcolor{red}{Red} boxes show the location of checks with work function 2.5~eV and
  \protect\textcolor{gray}{gray} boxes with 2.0~eV.}
  \label{fig:density_slice}
\end{figure}

Another quantity strongly affected by the work function distribution is the electron density. 
In \autoref{fig:density_slice}, we show a cross-sectional slice of the electron density
taken through the third row from the top in \autoref{fig:checkerboard_emittance}. 
Each curve was scaled such that the area under it is equal to~1. As a result, they show the distribution of particles emitted from different work function regions separately. If the density of electrons emitted from the areas of higher work function were scaled with respect to the one describing the electrons released from those of lower work function, then it would be so small that its shape would not be visible. 
The density of electrons emitted from lower work function areas forms maxima of a different height, which considerably decreases in magnitude toward the cathode center.
These electrons spread out above the areas of higher work function to such a degree that they overlap. Even greater spreading takes place to the vacuum outside the cathode and leads to the shift of the density peak associated with the edge patch away from the center of that check. The same features occur in the density of electrons emitted from the patches of higher work function but are much weaker, i.e., the density of electrons above the central areas is only slightly lower than above the edge one, the spreading of these beamlets is much smaller. The difference in the beamlets' qualitative structure from low and high work function is because space-charge effects are more pronounced for the areas with low work function, as the corresponding current density is much greater.

\begin{figure}
    \centering
    \includegraphics[]{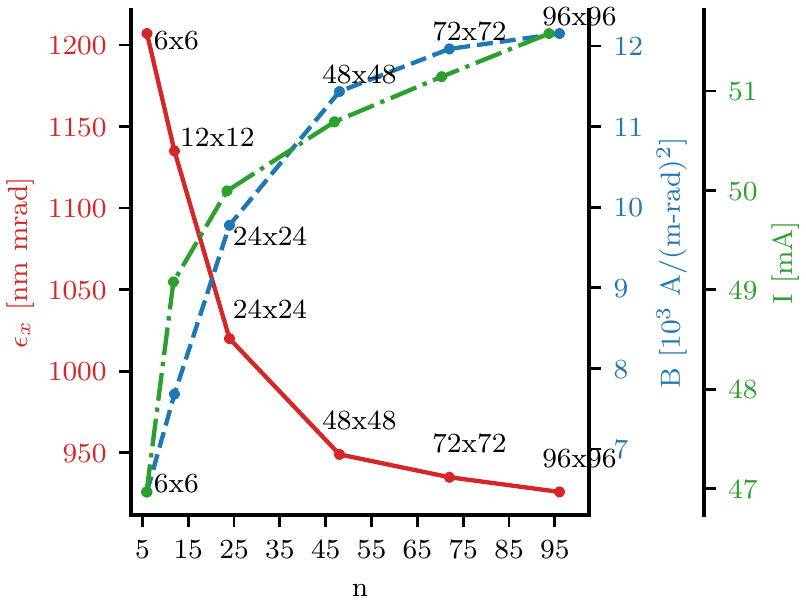}
    \caption{Emittance and brightness as a function of \(n\) of the checkerboard
    size \(n \times n\). The \textcolor{red}{red} solid line shows the emittance, the \textcolor{blue}{blue} dashed line the brightness, and the \textcolor{green}{green} dash dotted the current.}
    \label{fig:e_b_nxn}
  \end{figure}
The high edge emission due to decreased space charge effects on the sides of the uniform cathodes has been described previously~\cite{doi:10.1063/1.4978231,doi:10.1063/1.4914855}. In the case of checkerboard systems, this effect is seen not only at the outer edges of the emitter but also in each of the low work function checks so that the current density at the edge of those checks is greater than in their interior.
An increase of the number of checks results in an increase of emitted current as seen in \autoref{fig:e_b_nxn}, as the larger number of checks results in the increase of the total length of the edges, and thus the areas of the strongest emission. However, as the number of checks is increased while keeping the total emitter size unchanged, an increase in the number of checks corresponds to a decreased distance between checks of low work function. This leads to increased mutual space-charge effects between those checks, which has the effect of decreasing the emitted current from each of them~\cite{Haraldsson2020}. Eventually, the competing effects of enhanced edge emission and mutual space-charge interaction lead to the current leveling off at a value below that of a corresponding emitter of uniformly low work function.

In \autoref{fig:e_b_nxn}, we also see the \(x\) component of emittance's dependence on the number of checks on the emitter (the \(y\) component behaves similarly due to symmetry). The reason for the emittance decreasing with the increasing number of checks is the following.
When the checks are sufficiently large, they behave nearly like independent emitters, and the beam consists of small beamlets centered above low work function checks, distributed periodically over the entire emitter. Electrons emitted from the edge of each low work function check experience a transverse force due to higher space-charge above the check's interior than over the adjacent high work function check. As the checks become smaller and more numerous, the distance between the low work function checks decreases, and the asymmetry in the transverse space-charge force is reduced, which in turn reduces the transverse component of electrons emitted from the check edges. Thus, emittance decreases with increasing check number. A competing effect is that with greater current density (or simply greater electron density near the cathode), scattering increases, leading to greater emittance. Thus the growth in current with increasing the check number causes the emittance to grow. Hence, the emittance initially drops with increasing check number but then asymptotically reaches a constant value.
The decrease of emittance accompanied by the increase of the emitted current results in an increase of the brightness (\textcolor{blue}{blue} line \autoref{fig:e_b_nxn}). 
The finer the work function grid on the cathode's surface, the better quality of the emitted beam is obtained. It is interesting to note that the brightness of a fine-grained emitter, with alternating checks of high and low work function, can be considerably higher than that of an emitter with uniformly low work function, even with moderate reduction in current, as can be seen by comparison of \autoref{fig:e_b_nxn} with \autoref{fig:current_vs_mean_r} and \autoref{fig:brightness_data}.

\subsection{Random pattern}
\begin{figure}
  \centering
    {\phantomsubcaption\label{fig:random_pattern-a}}
    {\phantomsubcaption\label{fig:random_pattern-b}}
  \includegraphics[]{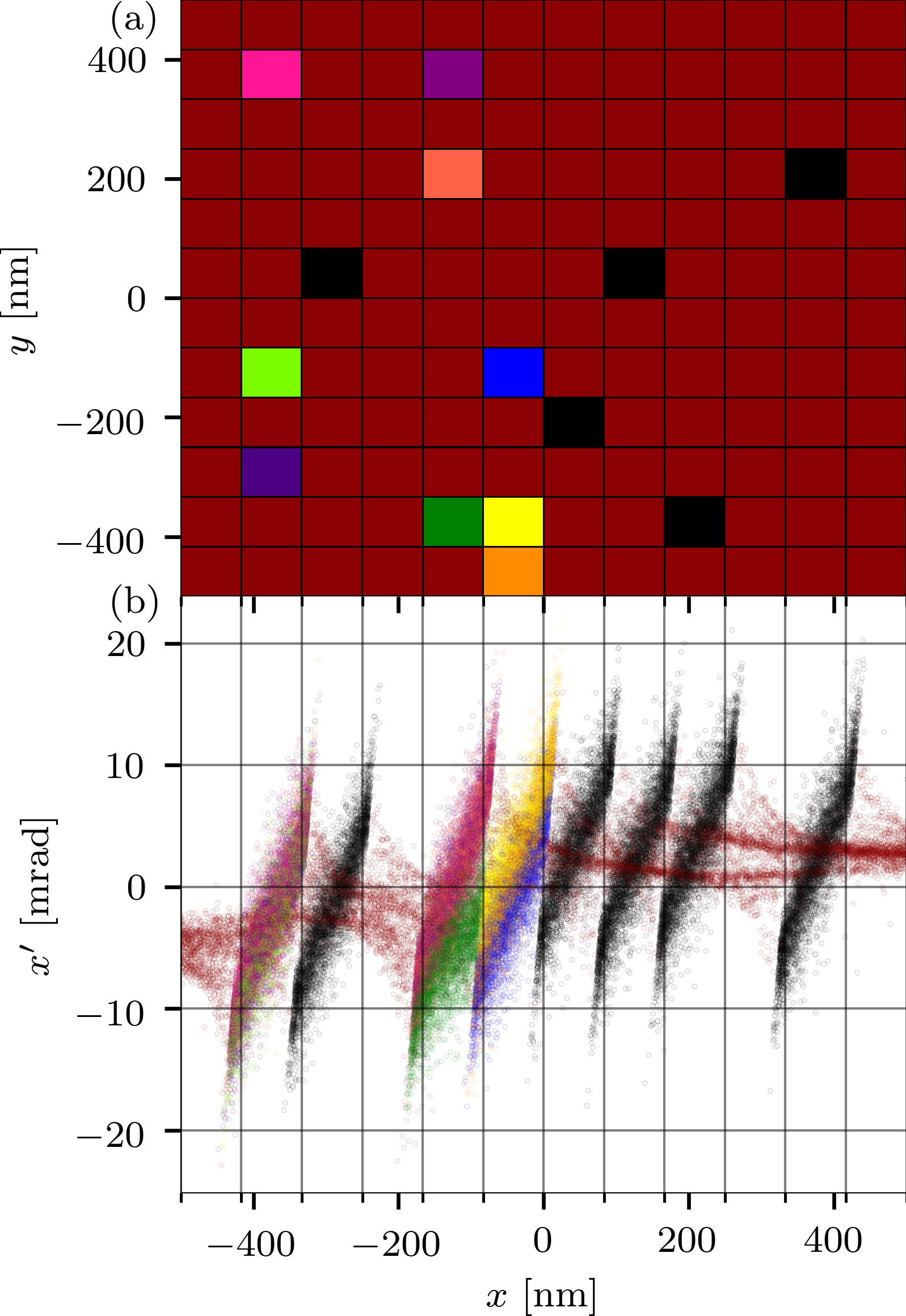}
  \caption{(a) A \(12\times 12\) lattice with work function 2.5~eV where 14 randomly distributed patches with a lower work function 2.0~eV
  have been placed on it. Patches with work function 2.5~eV are shown in \textcolor{red}{red} while patches in other colors correspond to the work function 2.0~eV.
  (b)~The \(x\)-\(x^\prime\) phase space diagram for the pattern shown in~a. The colors show which patch the electrons where emitted from.
  }
  \label{fig:random_pattern}
\end{figure}

We also examined a random distribution of patches on a square \(12\times 12\) lattice.
An example containing 14 patches of lower work function is shown in \autoref{fig:random_pattern-a}, and the corresponding phase-space diagram below in \autoref{fig:random_pattern-b}.
Every single area of lower work function induces a similar S-shaped pattern in the phase-space diagram (shown in the same color as the corresponding patch). The center and shape depend on the particular check position with respect to other low work function patches.  For instance, in the second column from the left of \autoref{fig:random_pattern-a}, there are three patches with low work function. The S-shaped patterns in phase space, due to electrons from these patches, mostly overlap as none of the low work function patches has a similar low work function patch adjacent to it in the first or third column. By comparison, in the fifth column from the left, there are two low work function patches. The upper one is surrounded by a high work function emitting area, whereas the lower (in green) one has two immediate neighbors of low work function, which affect it via space-charge forces. As a result, the green S-shape in phase space is skewed and shifted considerably to lower values of \(x^\prime\).

We performed simulations for 6, 14, 28, and 56 randomly distributed patches of either lower or higher work function on the same \(12\times 12\) lattice. For a fixed number of patches, 50 simulations were conducted, each with new random distribution and thus a different mean distance between the minority patches.

\begin{figure}
  \centering
    {\phantomsubcaption\label{fig:current_vs_mean_r-a}}
    {\phantomsubcaption\label{fig:current_vs_mean_r-b}}
  \includegraphics[]{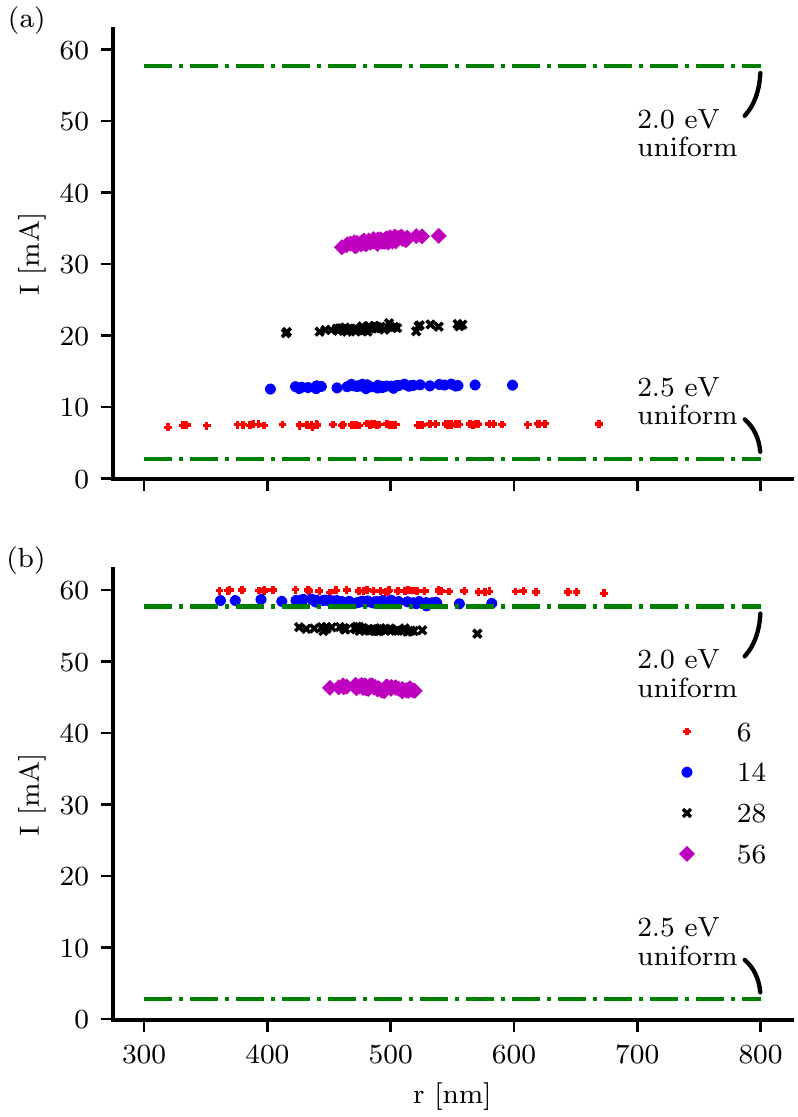}
  \caption{Current as a function of mean distance between minority work function patches.
  Simulations were done for 6 (\textcolor{red}{red \raisebox{0.2ex}{\(\scriptstyle +\)}}), 14 (\textcolor{blue}{blue \(\bullet\)}),
  28 (\textcolor{black}{\bf black \(\times\)}) and 56 (\textcolor{magenta}{magenta~\raisebox{0.2ex}{\(\scriptstyle\blacklozenge\)}})
  checks with work function different from the background
  randomly placed on the surface. The two \textcolor{green}{green} dash dotted lines show the two uniform cases 2.0~eV and 2.5~eV.
  (a) Low work function (\(\phi_{\rm low}\)) in minority.
  (b) High work function (\(\phi_{\rm high}\)) in minority.}
  \label{fig:current_vs_mean_r}
\end{figure}
\autoref{fig:current_vs_mean_r} shows the current as a function of the mean distance between all pairs of minority work function patches and the reference values of the current from the uniform cathodes with the two work function values used.
When the low work function patches are in the minority, the current is always between the values corresponding to uniform cathodes and strongly increases with the number of low work function areas.
The decrease of work function from \(2.5\) to \(2.0\)~eV results in a considerable increase of current from uniform cathodes and also from individual patches. As a result, the incorporation of even a few low work function patches against a high work function background increases the current.
Importantly, the increase in current is not proportional to the total area of low work function patches,
e.g., the introduction of 6 patches increases the current by about 2.7x. In comparison, the change from 28 to 56 patches increases the current by about 1.6x.
For a fixed number of minority checks the current increases slightly with the mean distance of added patches. The space-charge effects between patches decrease with increasing distance between low work function areas allowing them to emit more current. Moreover, the rate of increase becomes greater for a larger number of randomly distributed low work function checks. When there are only 6 low work function patches on the \(12\times 12\) emitter, then most arrangements are such that the low work function patches are essentially uncoupled, and the increase with the mean distance is very weak. The situation changes when the number of low work function patches increases because the areas may cluster together and be strongly coupled. This is seen for 56 patches when the lowest current is obtained when the minority checks are close to each other. The increase of the mean distance reduces the mutual space-charge effects between the low work function patches, leading to an increase of emission from them, thus increasing the total current.
An interesting effect occurs for the opposite case, i.e., when the high work function checks are the minority ones. A few such patches can increase the current above the level of a uniform cathode with the lower value of work function. As the number of high work function patches increases, the total current will drop down because the increase of current resulting from the reduction of space-charge effects becomes dominated by the decrease in current due to a larger fraction of the total area emitting low current.

\begin{figure}
  \centering
    {\phantomsubcaption\label{fig:emittance_data-a}}
    {\phantomsubcaption\label{fig:emittance_data-b}}
  \includegraphics[]{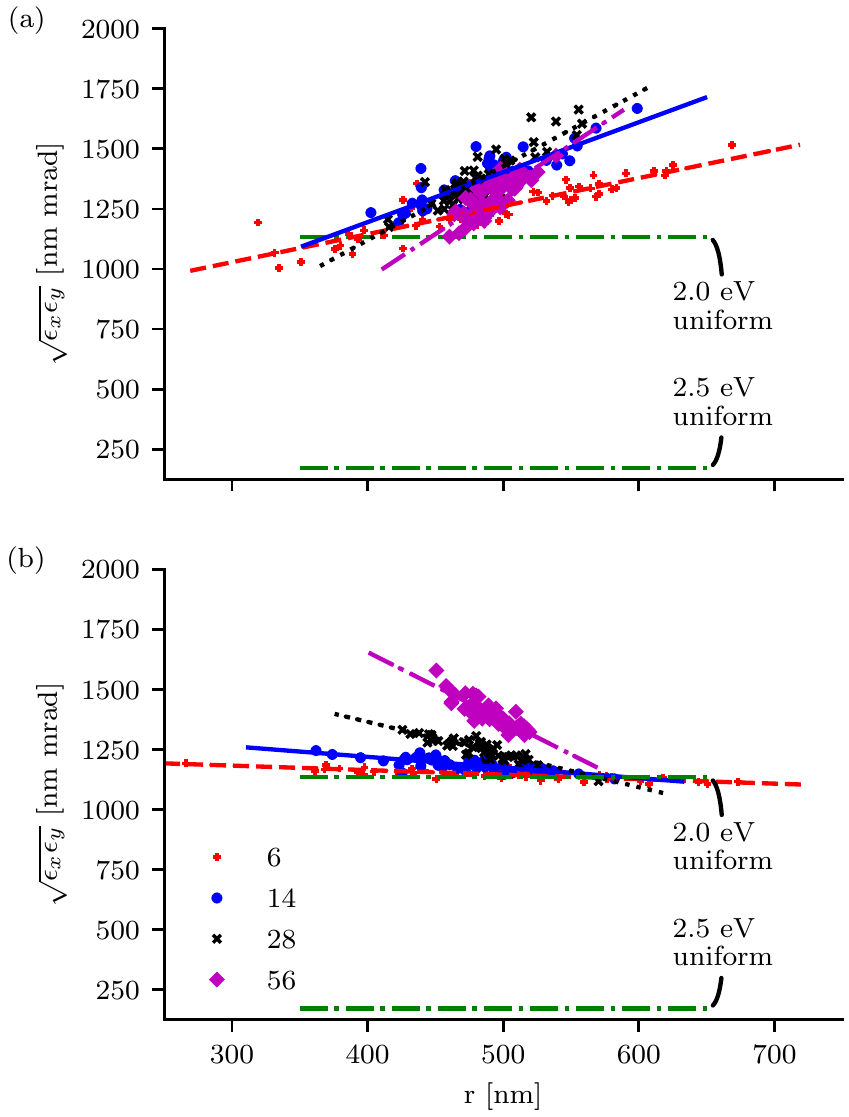}
  \caption{Emittance calculations for low/high work function patches that are randomly placed on a \(12\times 12\) board.
  Simulations were done for 6 (\textcolor{red}{red \raisebox{0.2ex}{\(\scriptstyle +\)}}), 14 (\textcolor{blue}{blue \(\bullet\)}),
  28 (\textcolor{black}{\bf black $\times$}) and 56 (\textcolor{magenta}{magenta \raisebox{0.2ex}{\(\scriptstyle\blacklozenge\)}}) checks with work function different from the background.
  The \textcolor{green}{green} dash dotted lines show the 2.0~eV and 2.5~eV uniform cases.
  (a) Low work function (\(\phi_{\rm low}\)) in minority.
  (b) High work function (\(\phi_{\rm high}\)) in minority.  }
  \label{fig:emittance_data}
\end{figure}
Contrary to the uniform and checkerboard systems, the emittance components in the \(x\) and \(y\) direction of the beam emitted from the cathodes containing randomly distributed minority patches are not the same. To capture the emittance for both directions in a single number, we multiply the two components and take their square root, \(\sqrt{\epsilon_x\epsilon_y}\). In \autoref{fig:emittance_data}, we plot this quantity as a function of the mean distance between all pairs of minority work function patches, and for reference, we show the emittance of the uniform cathodes. The emittance of the high work function uniform cathode is much lower than in the other cases, as is the current, \autoref{fig:current_vs_mean_r}. In this case, the beam consists of a few electrons separated from each other by distances at which the Coulomb repulsion is relatively weak, with minimal scattering. Thus their trajectories are mostly normal to the surface.      
Adding even a few low work function patches results in an increase of the emittance, which generally exceeds the low work function uniform cathode's emittance, \autoref{fig:emittance_data-a}. These checks provide the great majority of electrons in the beam. With greater spacing between the low work function patches, the beamlets have more room to expand transversely, which leads to greater emittance. Additionally, with greater spacing between low energy patches, the mutual space-charge coupling between them decreases, and the current density in the beamlets emanating from these patches is increased. With increased current density comes increased scattering, and in turn, increased emittance.
For the opposite case, i.e., when the high work function patches are the minority ones, their presence increases the emittance beyond the value for the low work function uniform cathode. Still, their impact decreases when their number is reduced and the separation increased, \autoref{fig:emittance_data-b}. 
Agglomerated patches of higher work function increase the beam's inhomogeneity. The effect is reduced when the additional checks are distributed more uniformly.

\begin{figure}
  \centering
    {\phantomsubcaption\label{fig:brightness_data-a}}
    {\phantomsubcaption\label{fig:brightness_data-b}}
  \includegraphics[]{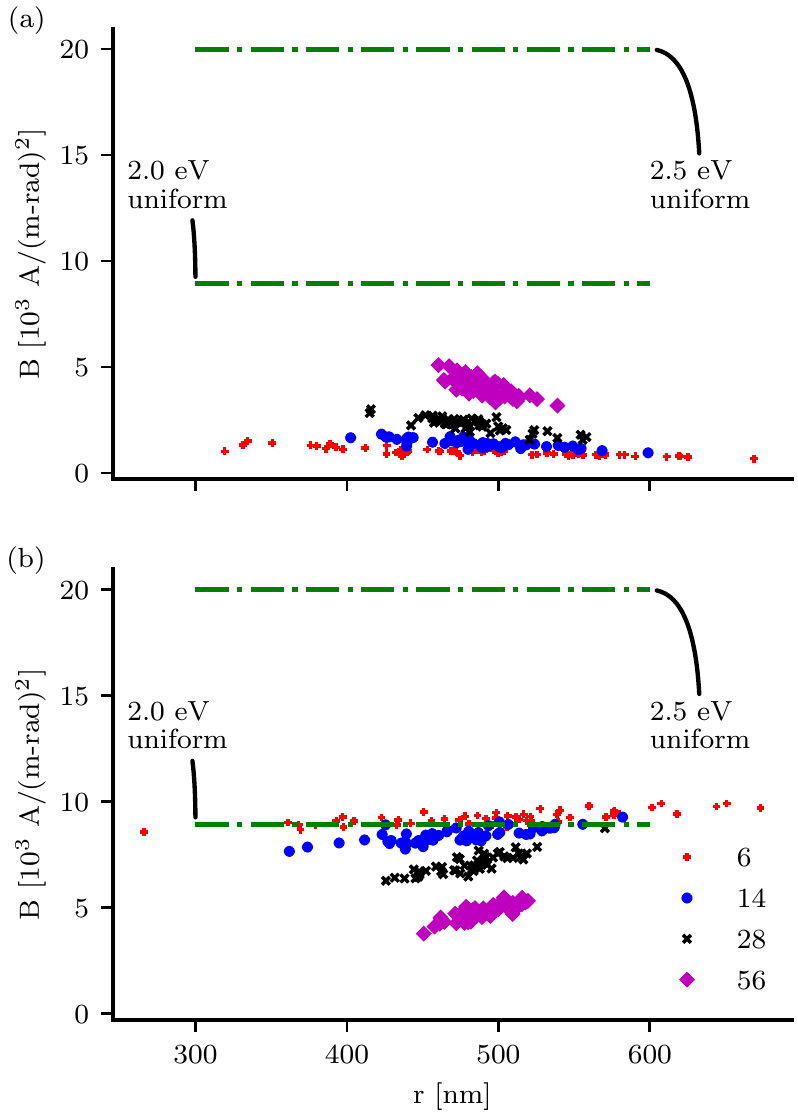}
  \caption{Brightness calculations for low/high work function patches that are randomly placed on a \(12\times 12\) board.
  Simulations were done for 6 (\textcolor{red}{red \raisebox{0.2ex}{\(\scriptstyle +\)}}), 14 (\textcolor{blue}{blue \(\bullet\)}),
  28 (\textcolor{black}{\bf black \(\times\)}) and 56 (\textcolor{magenta}{magenta \raisebox{0.2ex}{\(\scriptstyle\blacklozenge\)}})
 checks with work function different from the background.
  The \textcolor{green}{green} dash dotted line show the 2.0~eV and 2.5~eV uniform cases.
  (a) Low work function (\(\phi_{\rm low}\)) in minority.
  (b) High work function (\(\phi_{\rm high}\)) in minority.}
  \label{fig:brightness_data}
\end{figure}
Finally, in \autoref{fig:brightness_data}, we show the beam brightness versus the mean distance between pairs of minority work function patches. This quality factor is proportional to the ratio between the emitted current and the product of emittance components (\autoref{eq:brightness}). In the case of uniform systems, the very low
emittance of the high work function cathode leads to a greater brightness than for the cathode with the low work function, even though the current for the latter is an order of magnitude larger. 
Due to the sensitivity of the emittance to the introduction of low work function patches, as seen in \autoref{fig:emittance_data}, the low work function minority checks decrease the brightness below the value for the low work function uniform cathode, \autoref{fig:brightness_data-a}.
The much sensitivity of emittance to minority patch spacing [\autoref{fig:brightness_data}] compared to that of the current
[\autoref{fig:current_vs_mean_r}]
results in a decrease of the brightness with an increasing mean distance between the minority patches, \autoref{fig:brightness_data-a}.
The existence of minority checks with higher work function generally reduces the brightness with an increasing number of minority patches,
with the exception that a small number of such checks distributed over the surface may slightly increase the brightness beyond the value obtained for uniformly low work function emitter.
Due to the decrease of emittance observed in \autoref{fig:emittance_data-a}, the brightness increases with the mean distance between the minority patches, \autoref{fig:brightness_data-b}.

\section{Summary and conclusions\label{sec:summary}}
In this work, we examined the emittance and brightness from cathodes with a two level work function. We first studied a checkerboard with grid sizes ranging from 6x6 to 96x96, where the work function alternated between \(\phi_{\rm low} = 2.0\, \mathrm{eV}\) and \(\phi_{\rm high} = 2.5\, \mathrm{eV}\). It was found that as the grid becomes denser, the emittance decreases, and the brightness increases due to space-charge effects.

We also looked at a cathode with a \(12\times 12\) grid where we randomly decided the work function to be \(\phi_{\rm low}\) or \(\phi_{\rm high}\) in the grid.
First, we looked at the situation when the majority of the board had a high work function of \(\phi_{\rm high}\), and we randomly placed low work function, \(\phi_{\rm low}\), patches on it.
The number of low work function patches placed was either 6, 14, 28, or 56.
Then we examined the inverted case, i.e., was when the majority of the board had a low work function of 2.0~eV, and we randomly placed 6, 14, 28, or 56 high work function patches on the board.

The mean distance between all pairs of minority patches, i.e., either of low or of high work function type, was used to characterize each series of simulations. We found that the current increases slightly with the mean distance between the minority patches when they have low work function $\phi_{\rm low}$, and it decreases slightly when they have high work function $\phi_{\rm high}$. The emittance showed similar behavior but much more pronounced than the current. The higher sensitivity of the emittance compared to the current is in line with the previous work by Jensen et~al.~\cite{doi:10.1116/1.2827508}. Consequently, the beam brightness is also sensitive, but with the opposite behavior, i.e., decreasing with the mean distance between the low work function patches when they were in the minority and increasing when the high work function patches were in the minority.

However, overall, the cathode's emittance with non-uniform work function is greater than for the uniform case, and at the same time the brightness decreases. The emittance results from lateral motion of the particles due to both scattering and irregular distribution of space-charge over the cathode surface, i.e., non-uniform beam divergence. In the homogeneous case, the electron density profile in the vicinity of the cathode is nearly uniform, except at the boundary regions. The work function variation leads to an inhomogeneous density profile, with zones of higher density emerging from the cathode spots with low work function, which tend to diffuse into the zones of lower density emerging from the spots with high work function. These imbalanced lateral forces lead to an extra divergence of the beam and thus to an increased emittance. 

\begin{acknowledgments}
This material is based upon work supported by the Air Force Office of Scientific Research under award number FA9550-18-1-7011, and by the Icelandic Research Fund grant number 174127-051.
Any opinions, findings, and conclusions or recommendations expressed in this material are those of the author and do not necessarily reflect the views of the United States Air Force.
\end{acknowledgments}

\bibliography{bibliography}

\begin{thebibliography}{24}%
\makeatletter
\providecommand \@ifxundefined [1]{%
 \@ifx{#1\undefined}
}%
\providecommand \@ifnum [1]{%
 \ifnum #1\expandafter \@firstoftwo
 \else \expandafter \@secondoftwo
 \fi
}%
\providecommand \@ifx [1]{%
 \ifx #1\expandafter \@firstoftwo
 \else \expandafter \@secondoftwo
 \fi
}%
\providecommand \natexlab [1]{#1}%
\providecommand \enquote  [1]{``#1''}%
\providecommand \bibnamefont  [1]{#1}%
\providecommand \bibfnamefont [1]{#1}%
\providecommand \citenamefont [1]{#1}%
\providecommand \href@noop [0]{\@secondoftwo}%
\providecommand \href [0]{\begingroup \@sanitize@url \@href}%
\providecommand \@href[1]{\@@startlink{#1}\@@href}%
\providecommand \@@href[1]{\endgroup#1\@@endlink}%
\providecommand \@sanitize@url [0]{\catcode `\\12\catcode `\$12\catcode
  `\&12\catcode `\#12\catcode `\^12\catcode `\_12\catcode `\%12\relax}%
\providecommand \@@startlink[1]{}%
\providecommand \@@endlink[0]{}%
\providecommand \url  [0]{\begingroup\@sanitize@url \@url }%
\providecommand \@url [1]{\endgroup\@href {#1}{\urlprefix }}%
\providecommand \urlprefix  [0]{URL }%
\providecommand \Eprint [0]{\href }%
\providecommand \doibase [0]{https://doi.org/}%
\providecommand \selectlanguage [0]{\@gobble}%
\providecommand \bibinfo  [0]{\@secondoftwo}%
\providecommand \bibfield  [0]{\@secondoftwo}%
\providecommand \translation [1]{[#1]}%
\providecommand \BibitemOpen [0]{}%
\providecommand \bibitemStop [0]{}%
\providecommand \bibitemNoStop [0]{.\EOS\space}%
\providecommand \EOS [0]{\spacefactor3000\relax}%
\providecommand \BibitemShut  [1]{\csname bibitem#1\endcsname}%
\let\auto@bib@innerbib\@empty
\bibitem [{\citenamefont {Torfason}\ \emph {et~al.}(2015)\citenamefont
  {Torfason}, \citenamefont {Valfells},\ and\ \citenamefont
  {Manolescu}}]{doi:10.1063/1.4914855}%
  \BibitemOpen
  \bibfield  {author} {\bibinfo {author} {\bibfnamefont {K.}~\bibnamefont
  {Torfason}}, \bibinfo {author} {\bibfnamefont {A.}~\bibnamefont {Valfells}},\
  and\ \bibinfo {author} {\bibfnamefont {A.}~\bibnamefont {Manolescu}},\
  }\bibfield  {title} {\bibinfo {title} {Molecular dynamics simulations of
  field emission from a planar nanodiode},\ }\href
  {https://doi.org/10.1063/1.4914855} {\bibfield  {journal} {\bibinfo
  {journal} {Physics of Plasmas}\ }\textbf {\bibinfo {volume} {22}},\ \bibinfo
  {pages} {033109} (\bibinfo {year} {2015})},\ \Eprint
  {https://arxiv.org/abs/https://doi.org/10.1063/1.4914855}
  {https://doi.org/10.1063/1.4914855} \BibitemShut {NoStop}%
\bibitem [{\citenamefont {Torfason}\ \emph {et~al.}(2016)\citenamefont
  {Torfason}, \citenamefont {Valfells},\ and\ \citenamefont
  {Manolescu}}]{doi:10.1063/1.4972821}%
  \BibitemOpen
  \bibfield  {author} {\bibinfo {author} {\bibfnamefont {K.}~\bibnamefont
  {Torfason}}, \bibinfo {author} {\bibfnamefont {A.}~\bibnamefont {Valfells}},\
  and\ \bibinfo {author} {\bibfnamefont {A.}~\bibnamefont {Manolescu}},\
  }\bibfield  {title} {\bibinfo {title} {Molecular dynamics simulations of
  field emission from a prolate spheroidal tip},\ }\href
  {https://doi.org/10.1063/1.4972821} {\bibfield  {journal} {\bibinfo
  {journal} {Physics of Plasmas}\ }\textbf {\bibinfo {volume} {23}},\ \bibinfo
  {pages} {123119} (\bibinfo {year} {2016})},\ \Eprint
  {https://arxiv.org/abs/https://doi.org/10.1063/1.4972821}
  {https://doi.org/10.1063/1.4972821} \BibitemShut {NoStop}%
\bibitem [{\citenamefont {Stoner}\ and\ \citenamefont
  {Glass}(2012)}]{stoner2012nanoelectronics}%
  \BibitemOpen
  \bibfield  {author} {\bibinfo {author} {\bibfnamefont {B.~R.}\ \bibnamefont
  {Stoner}}\ and\ \bibinfo {author} {\bibfnamefont {J.~T.}\ \bibnamefont
  {Glass}},\ }\bibfield  {title} {\bibinfo {title} {Nothing is like a vacuum},\
  }\href {https://doi.org/10.1038/nnano.2012.130} {\bibfield  {journal}
  {\bibinfo  {journal} {Nature Nanotechnology}\ }\textbf {\bibinfo {volume}
  {7}},\ \bibinfo {pages} {485} (\bibinfo {year} {2012})}\BibitemShut {NoStop}%
\bibitem [{\citenamefont {Han}\ \emph {et~al.}(2012)\citenamefont {Han},
  \citenamefont {Sub~Oh},\ and\ \citenamefont
  {Meyyappan}}]{doi:10.1063/1.4717751}%
  \BibitemOpen
  \bibfield  {author} {\bibinfo {author} {\bibfnamefont {J.-W.}\ \bibnamefont
  {Han}}, \bibinfo {author} {\bibfnamefont {J.}~\bibnamefont {Sub~Oh}},\ and\
  \bibinfo {author} {\bibfnamefont {M.}~\bibnamefont {Meyyappan}},\ }\bibfield
  {title} {\bibinfo {title} {Vacuum nanoelectronics: Back to the future?—gate
  insulated nanoscale vacuum channel transistor},\ }\href
  {https://doi.org/10.1063/1.4717751} {\bibfield  {journal} {\bibinfo
  {journal} {Applied Physics Letters}\ }\textbf {\bibinfo {volume} {100}},\
  \bibinfo {pages} {213505} (\bibinfo {year} {2012})},\ \Eprint
  {https://arxiv.org/abs/https://doi.org/10.1063/1.4717751}
  {https://doi.org/10.1063/1.4717751} \BibitemShut {NoStop}%
\bibitem [{\citenamefont {Zhang}\ and\ \citenamefont
  {Lau}(2016)}]{zhang2016ultrafast}%
  \BibitemOpen
  \bibfield  {author} {\bibinfo {author} {\bibfnamefont {P.}~\bibnamefont
  {Zhang}}\ and\ \bibinfo {author} {\bibfnamefont {Y.}~\bibnamefont {Lau}},\
  }\bibfield  {title} {\bibinfo {title} {Ultrafast strong-field photoelectron
  emission from biased metal surfaces: exact solution to time-dependent
  schr{\"o}dinger equation},\ }\href {https://doi.org/10.1038/srep19894}
  {\bibfield  {journal} {\bibinfo  {journal} {Scientific reports}\ }\textbf
  {\bibinfo {volume} {6}},\ \bibinfo {pages} {19894} (\bibinfo {year}
  {2016})}\BibitemShut {NoStop}%
\bibitem [{\citenamefont {Dombi}\ \emph {et~al.}(2020)\citenamefont {Dombi},
  \citenamefont {P\'apa}, \citenamefont {Vogelsang}, \citenamefont {Yalunin},
  \citenamefont {Sivis}, \citenamefont {Herink}, \citenamefont {Sch\"afer},
  \citenamefont {Gro\ss{}}, \citenamefont {Ropers},\ and\ \citenamefont
  {Lienau}}]{RevModPhys.92.025003}%
  \BibitemOpen
  \bibfield  {author} {\bibinfo {author} {\bibfnamefont {P.}~\bibnamefont
  {Dombi}}, \bibinfo {author} {\bibfnamefont {Z.}~\bibnamefont {P\'apa}},
  \bibinfo {author} {\bibfnamefont {J.}~\bibnamefont {Vogelsang}}, \bibinfo
  {author} {\bibfnamefont {S.~V.}\ \bibnamefont {Yalunin}}, \bibinfo {author}
  {\bibfnamefont {M.}~\bibnamefont {Sivis}}, \bibinfo {author} {\bibfnamefont
  {G.}~\bibnamefont {Herink}}, \bibinfo {author} {\bibfnamefont
  {S.}~\bibnamefont {Sch\"afer}}, \bibinfo {author} {\bibfnamefont
  {P.}~\bibnamefont {Gro\ss{}}}, \bibinfo {author} {\bibfnamefont
  {C.}~\bibnamefont {Ropers}},\ and\ \bibinfo {author} {\bibfnamefont
  {C.}~\bibnamefont {Lienau}},\ }\bibfield  {title} {\bibinfo {title}
  {Strong-field nano-optics},\ }\href
  {https://doi.org/10.1103/RevModPhys.92.025003} {\bibfield  {journal}
  {\bibinfo  {journal} {Rev. Mod. Phys.}\ }\textbf {\bibinfo {volume} {92}},\
  \bibinfo {pages} {025003} (\bibinfo {year} {2020})}\BibitemShut {NoStop}%
\bibitem [{\citenamefont {Feist}\ \emph {et~al.}(2017)\citenamefont {Feist},
  \citenamefont {Bach}, \citenamefont {{Rubiano da Silva}}, \citenamefont
  {Danz}, \citenamefont {Möller}, \citenamefont {Priebe}, \citenamefont
  {Domröse}, \citenamefont {Gatzmann}, \citenamefont {Rost}, \citenamefont
  {Schauss}, \citenamefont {Strauch}, \citenamefont {Bormann}, \citenamefont
  {Sivis}, \citenamefont {Schäfer},\ and\ \citenamefont
  {Ropers}}]{FEIST201763}%
  \BibitemOpen
  \bibfield  {author} {\bibinfo {author} {\bibfnamefont {A.}~\bibnamefont
  {Feist}}, \bibinfo {author} {\bibfnamefont {N.}~\bibnamefont {Bach}},
  \bibinfo {author} {\bibfnamefont {N.}~\bibnamefont {{Rubiano da Silva}}},
  \bibinfo {author} {\bibfnamefont {T.}~\bibnamefont {Danz}}, \bibinfo {author}
  {\bibfnamefont {M.}~\bibnamefont {Möller}}, \bibinfo {author} {\bibfnamefont
  {K.~E.}\ \bibnamefont {Priebe}}, \bibinfo {author} {\bibfnamefont
  {T.}~\bibnamefont {Domröse}}, \bibinfo {author} {\bibfnamefont {J.~G.}\
  \bibnamefont {Gatzmann}}, \bibinfo {author} {\bibfnamefont {S.}~\bibnamefont
  {Rost}}, \bibinfo {author} {\bibfnamefont {J.}~\bibnamefont {Schauss}},
  \bibinfo {author} {\bibfnamefont {S.}~\bibnamefont {Strauch}}, \bibinfo
  {author} {\bibfnamefont {R.}~\bibnamefont {Bormann}}, \bibinfo {author}
  {\bibfnamefont {M.}~\bibnamefont {Sivis}}, \bibinfo {author} {\bibfnamefont
  {S.}~\bibnamefont {Schäfer}},\ and\ \bibinfo {author} {\bibfnamefont
  {C.}~\bibnamefont {Ropers}},\ }\bibfield  {title} {\bibinfo {title}
  {Ultrafast transmission electron microscopy using a laser-driven field
  emitter: Femtosecond resolution with a high coherence electron beam},\ }\href
  {https://doi.org/https://doi.org/10.1016/j.ultramic.2016.12.005} {\bibfield
  {journal} {\bibinfo  {journal} {Ultramicroscopy}\ }\textbf {\bibinfo {volume}
  {176}},\ \bibinfo {pages} {63 } (\bibinfo {year} {2017})},\ \bibinfo {note}
  {70th Birthday of Robert Sinclair and 65th Birthday of Nestor J. Zaluzec PICO
  2017 – Fourth Conference on Frontiers of Aberration Corrected Electron
  Microscopy}\BibitemShut {NoStop}%
\bibitem [{\citenamefont {{Marrese}}\ \emph {et~al.}(2000)\citenamefont
  {{Marrese}}, \citenamefont {{Polk}}, \citenamefont {{Jensen}}, \citenamefont
  {{Gallimore}}, \citenamefont {{Spindt}}, \citenamefont {{Fink}},\ and\
  \citenamefont {{Devereux Palmer}}}]{2000mfss.conf..271M}%
  \BibitemOpen
  \bibfield  {author} {\bibinfo {author} {\bibfnamefont {C.~M.}\ \bibnamefont
  {{Marrese}}}, \bibinfo {author} {\bibfnamefont {J.~E.}\ \bibnamefont
  {{Polk}}}, \bibinfo {author} {\bibfnamefont {K.~L.}\ \bibnamefont
  {{Jensen}}}, \bibinfo {author} {\bibfnamefont {A.~D.}\ \bibnamefont
  {{Gallimore}}}, \bibinfo {author} {\bibfnamefont {C.~A.}\ \bibnamefont
  {{Spindt}}}, \bibinfo {author} {\bibfnamefont {R.~L.}\ \bibnamefont
  {{Fink}}},\ and\ \bibinfo {author} {\bibfnamefont {W.}~\bibnamefont
  {{Devereux Palmer}}},\ }\bibfield  {title} {\bibinfo {title} {{Performance of
  Field Emission Cathodes in Xenon Electric Propulsion System Environments}},\
  }in\ \href@noop {} {\emph {\bibinfo {booktitle} {Micropropulsion for Small
  Spacecraft}}},\ \bibinfo {editor} {edited by\ \bibinfo {editor}
  {\bibfnamefont {M.~M.}\ \bibnamefont {{Micci}}}\ and\ \bibinfo {editor}
  {\bibfnamefont {A.~D.}\ \bibnamefont {{Ketsdever}}}}\ (\bibinfo {year}
  {2000})\ p.\ \bibinfo {pages} {271}\BibitemShut {NoStop}%
\bibitem [{\citenamefont {Jensen}\ \emph {et~al.}(2014)\citenamefont {Jensen},
  \citenamefont {Shiffler}, \citenamefont {Petillo}, \citenamefont {Pan},\ and\
  \citenamefont {Luginsland}}]{PhysRevSTAB.17.043402}%
  \BibitemOpen
  \bibfield  {author} {\bibinfo {author} {\bibfnamefont {K.~L.}\ \bibnamefont
  {Jensen}}, \bibinfo {author} {\bibfnamefont {D.~A.}\ \bibnamefont
  {Shiffler}}, \bibinfo {author} {\bibfnamefont {J.~J.}\ \bibnamefont
  {Petillo}}, \bibinfo {author} {\bibfnamefont {Z.}~\bibnamefont {Pan}},\ and\
  \bibinfo {author} {\bibfnamefont {J.~W.}\ \bibnamefont {Luginsland}},\
  }\bibfield  {title} {\bibinfo {title} {Emittance, surface structure, and
  electron emission},\ }\href {https://doi.org/10.1103/PhysRevSTAB.17.043402}
  {\bibfield  {journal} {\bibinfo  {journal} {Phys. Rev. ST Accel. Beams}\
  }\textbf {\bibinfo {volume} {17}},\ \bibinfo {pages} {043402} (\bibinfo
  {year} {2014})}\BibitemShut {NoStop}%
\bibitem [{\citenamefont {Jensen}\ \emph {et~al.}(2019)\citenamefont {Jensen},
  \citenamefont {McDonald}, \citenamefont {Chubenko}, \citenamefont {Harris},
  \citenamefont {Shiffler}, \citenamefont {Moody}, \citenamefont {Petillo},\
  and\ \citenamefont {Jensen}}]{doi:10.1063/1.5097149}%
  \BibitemOpen
  \bibfield  {author} {\bibinfo {author} {\bibfnamefont {K.~L.}\ \bibnamefont
  {Jensen}}, \bibinfo {author} {\bibfnamefont {M.}~\bibnamefont {McDonald}},
  \bibinfo {author} {\bibfnamefont {O.}~\bibnamefont {Chubenko}}, \bibinfo
  {author} {\bibfnamefont {J.~R.}\ \bibnamefont {Harris}}, \bibinfo {author}
  {\bibfnamefont {D.~A.}\ \bibnamefont {Shiffler}}, \bibinfo {author}
  {\bibfnamefont {N.~A.}\ \bibnamefont {Moody}}, \bibinfo {author}
  {\bibfnamefont {J.~J.}\ \bibnamefont {Petillo}},\ and\ \bibinfo {author}
  {\bibfnamefont {A.~J.}\ \bibnamefont {Jensen}},\ }\bibfield  {title}
  {\bibinfo {title} {Thermal-field and photoemission from meso- and micro-scale
  features: Effects of screening and roughness on characterization and
  simulation},\ }\href {https://doi.org/10.1063/1.5097149} {\bibfield
  {journal} {\bibinfo  {journal} {Journal of Applied Physics}\ }\textbf
  {\bibinfo {volume} {125}},\ \bibinfo {pages} {234303} (\bibinfo {year}
  {2019})},\ \Eprint {https://arxiv.org/abs/https://doi.org/10.1063/1.5097149}
  {https://doi.org/10.1063/1.5097149} \BibitemShut {NoStop}%
\bibitem [{\citenamefont {Lau}(1987)}]{doi:10.1063/1.338833}%
  \BibitemOpen
  \bibfield  {author} {\bibinfo {author} {\bibfnamefont {Y.~Y.}\ \bibnamefont
  {Lau}},\ }\bibfield  {title} {\bibinfo {title} {Effects of cathode surface
  roughness on the quality of electron beams},\ }\href
  {https://doi.org/10.1063/1.338833} {\bibfield  {journal} {\bibinfo  {journal}
  {Journal of Applied Physics}\ }\textbf {\bibinfo {volume} {61}},\ \bibinfo
  {pages} {36} (\bibinfo {year} {1987})},\ \Eprint
  {https://arxiv.org/abs/https://doi.org/10.1063/1.338833}
  {https://doi.org/10.1063/1.338833} \BibitemShut {NoStop}%
\bibitem [{\citenamefont {Krasilnikov}(2006)}]{krasilnikov2006impact}%
  \BibitemOpen
  \bibfield  {author} {\bibinfo {author} {\bibfnamefont {M.}~\bibnamefont
  {Krasilnikov}},\ }\bibfield  {title} {\bibinfo {title} {Impact of the cathode
  roughness on the emittance of an electron beam},\ }in\ \href@noop {} {\emph
  {\bibinfo {booktitle} {Proceedings of FEL 2006 Conference, Berlin,
  Germany}}}\ (\bibinfo {year} {2006})\ pp.\ \bibinfo {pages}
  {583--586}\BibitemShut {NoStop}%
\bibitem [{\citenamefont {Ilkov}\ \emph {et~al.}(2015)\citenamefont {Ilkov},
  \citenamefont {Torfason}, \citenamefont {Manolescu},\ and\ \citenamefont
  {Valfells}}]{Ilkov2015}%
  \BibitemOpen
  \bibfield  {author} {\bibinfo {author} {\bibfnamefont {M.}~\bibnamefont
  {Ilkov}}, \bibinfo {author} {\bibfnamefont {K.}~\bibnamefont {Torfason}},
  \bibinfo {author} {\bibfnamefont {A.}~\bibnamefont {Manolescu}},\ and\
  \bibinfo {author} {\bibfnamefont {A.}~\bibnamefont {Valfells}},\ }\bibfield
  {title} {\bibinfo {title} {Terahertz pulsed photogenerated current in
  microdiodes at room temperature},\ }\href {https://doi.org/10.1063/1.4936176}
  {\bibfield  {journal} {\bibinfo  {journal} {Applied Physics Letters}\
  }\textbf {\bibinfo {volume} {107}},\ \bibinfo {pages} {203508} (\bibinfo
  {year} {2015})},\ \Eprint
  {https://arxiv.org/abs/https://doi.org/10.1063/1.4936176}
  {https://doi.org/10.1063/1.4936176} \BibitemShut {NoStop}%
\bibitem [{\citenamefont {Jensen}\ \emph {et~al.}(2008)\citenamefont {Jensen},
  \citenamefont {Petillo}, \citenamefont {Montgomery}, \citenamefont {Pan},
  \citenamefont {Feldman}, \citenamefont {O’Shea}, \citenamefont {Moody},
  \citenamefont {Cahay}, \citenamefont {Yater},\ and\ \citenamefont
  {Shaw}}]{doi:10.1116/1.2827508}%
  \BibitemOpen
  \bibfield  {author} {\bibinfo {author} {\bibfnamefont {K.~L.}\ \bibnamefont
  {Jensen}}, \bibinfo {author} {\bibfnamefont {J.~J.}\ \bibnamefont {Petillo}},
  \bibinfo {author} {\bibfnamefont {E.~J.}\ \bibnamefont {Montgomery}},
  \bibinfo {author} {\bibfnamefont {Z.}~\bibnamefont {Pan}}, \bibinfo {author}
  {\bibfnamefont {D.~W.}\ \bibnamefont {Feldman}}, \bibinfo {author}
  {\bibfnamefont {P.~G.}\ \bibnamefont {O’Shea}}, \bibinfo {author}
  {\bibfnamefont {N.~A.}\ \bibnamefont {Moody}}, \bibinfo {author}
  {\bibfnamefont {M.}~\bibnamefont {Cahay}}, \bibinfo {author} {\bibfnamefont
  {J.~E.}\ \bibnamefont {Yater}},\ and\ \bibinfo {author} {\bibfnamefont
  {J.~L.}\ \bibnamefont {Shaw}},\ }\bibfield  {title} {\bibinfo {title}
  {Application of a general electron emission equation to surface nonuniformity
  and current density variation},\ }\href {https://doi.org/10.1116/1.2827508}
  {\bibfield  {journal} {\bibinfo  {journal} {Journal of Vacuum Science \&
  Technology B: Microelectronics and Nanometer Structures Processing,
  Measurement, and Phenomena}\ }\textbf {\bibinfo {volume} {26}},\ \bibinfo
  {pages} {831} (\bibinfo {year} {2008})},\ \Eprint
  {https://arxiv.org/abs/https://avs.scitation.org/doi/pdf/10.1116/1.2827508}
  {https://avs.scitation.org/doi/pdf/10.1116/1.2827508} \BibitemShut {NoStop}%
\bibitem [{\citenamefont {O{\textquoteright}Shea}\ and\ \citenamefont
  {Freund}(2001)}]{OShea1853}%
  \BibitemOpen
  \bibfield  {author} {\bibinfo {author} {\bibfnamefont {P.~G.}\ \bibnamefont
  {O{\textquoteright}Shea}}\ and\ \bibinfo {author} {\bibfnamefont {H.~P.}\
  \bibnamefont {Freund}},\ }\bibfield  {title} {\bibinfo {title} {Free-electron
  lasers: Status and applications},\ }\href
  {https://doi.org/10.1126/science.1055718} {\bibfield  {journal} {\bibinfo
  {journal} {Science}\ }\textbf {\bibinfo {volume} {292}},\ \bibinfo {pages}
  {1853} (\bibinfo {year} {2001})},\ \Eprint
  {https://arxiv.org/abs/https://science.sciencemag.org/content/292/5523/1853.full.pdf}
  {https://science.sciencemag.org/content/292/5523/1853.full.pdf} \BibitemShut
  {NoStop}%
\bibitem [{\citenamefont {{Parker}}\ \emph {et~al.}(2002)\citenamefont
  {{Parker}}, \citenamefont {{Abrams}}, \citenamefont {{Danly}},\ and\
  \citenamefont {{Levush}}}]{989967}%
  \BibitemOpen
  \bibfield  {author} {\bibinfo {author} {\bibfnamefont {R.~K.}\ \bibnamefont
  {{Parker}}}, \bibinfo {author} {\bibfnamefont {R.~H.}\ \bibnamefont
  {{Abrams}}}, \bibinfo {author} {\bibfnamefont {B.~G.}\ \bibnamefont
  {{Danly}}},\ and\ \bibinfo {author} {\bibfnamefont {B.}~\bibnamefont
  {{Levush}}},\ }\bibfield  {title} {\bibinfo {title} {Vacuum electronics},\
  }\href@noop {} {\bibfield  {journal} {\bibinfo  {journal} {IEEE Transactions
  on Microwave Theory and Techniques}\ }\textbf {\bibinfo {volume} {50}},\
  \bibinfo {pages} {835} (\bibinfo {year} {2002})}\BibitemShut {NoStop}%
\bibitem [{\citenamefont {{Haraldsson}}\ \emph {et~al.}(2020)\citenamefont
  {{Haraldsson}}, \citenamefont {{Torfason}}, \citenamefont {{Manolescu}},\
  and\ \citenamefont {{Valfells}}}]{Haraldsson2020}%
  \BibitemOpen
  \bibfield  {author} {\bibinfo {author} {\bibfnamefont {H.~V.}\ \bibnamefont
  {{Haraldsson}}}, \bibinfo {author} {\bibfnamefont {K.}~\bibnamefont
  {{Torfason}}}, \bibinfo {author} {\bibfnamefont {A.}~\bibnamefont
  {{Manolescu}}},\ and\ \bibinfo {author} {\bibfnamefont {A.}~\bibnamefont
  {{Valfells}}},\ }\bibfield  {title} {\bibinfo {title} {Molecular dynamics
  simulations of mutual space-charge effect between planar field emitters},\
  }\href {https://doi.org/10.1109/TPS.2020.2991582} {\bibfield  {journal}
  {\bibinfo  {journal} {IEEE Transactions on Plasma Science}\ }\textbf
  {\bibinfo {volume} {48}},\ \bibinfo {pages} {1967} (\bibinfo {year}
  {2020})}\BibitemShut {NoStop}%
\bibitem [{\citenamefont {Forbes}\ and\ \citenamefont
  {Deane}(2007)}]{Forbes08112007}%
  \BibitemOpen
  \bibfield  {author} {\bibinfo {author} {\bibfnamefont {R.~G.}\ \bibnamefont
  {Forbes}}\ and\ \bibinfo {author} {\bibfnamefont {J.~H.}\ \bibnamefont
  {Deane}},\ }\bibfield  {title} {\bibinfo {title} {Reformulation of the
  standard theory of fowler–nordheim tunnelling and cold field electron
  emission},\ }\href {https://doi.org/10.1098/rspa.2007.0030} {\bibfield
  {journal} {\bibinfo  {journal} {Proceedings of the Royal Society A:
  Mathematical, Physical and Engineering Science}\ }\textbf {\bibinfo {volume}
  {463}},\ \bibinfo {pages} {2907} (\bibinfo {year} {2007})}\BibitemShut
  {NoStop}%
\bibitem [{\citenamefont {Hahn}(2005)}]{HAHN200578}%
  \BibitemOpen
  \bibfield  {author} {\bibinfo {author} {\bibfnamefont {T.}~\bibnamefont
  {Hahn}},\ }\bibfield  {title} {\bibinfo {title} {Cuba—a library for
  multidimensional numerical integration},\ }\href
  {https://doi.org/https://doi.org/10.1016/j.cpc.2005.01.010} {\bibfield
  {journal} {\bibinfo  {journal} {Computer Physics Communications}\ }\textbf
  {\bibinfo {volume} {168}},\ \bibinfo {pages} {78 } (\bibinfo {year}
  {2005})}\BibitemShut {NoStop}%
\bibitem [{\citenamefont {Shockley}(1938)}]{doi:10.1063/1.1710367}%
  \BibitemOpen
  \bibfield  {author} {\bibinfo {author} {\bibfnamefont {W.}~\bibnamefont
  {Shockley}},\ }\bibfield  {title} {\bibinfo {title} {Currents to conductors
  induced by a moving point charge},\ }\href
  {https://doi.org/10.1063/1.1710367} {\bibfield  {journal} {\bibinfo
  {journal} {Journal of Applied Physics}\ }\textbf {\bibinfo {volume} {9}},\
  \bibinfo {pages} {635} (\bibinfo {year} {1938})},\ \Eprint
  {https://arxiv.org/abs/https://doi.org/10.1063/1.1710367}
  {https://doi.org/10.1063/1.1710367} \BibitemShut {NoStop}%
\bibitem [{\citenamefont {{Ramo}}(1939)}]{1686997}%
  \BibitemOpen
  \bibfield  {author} {\bibinfo {author} {\bibfnamefont {S.}~\bibnamefont
  {{Ramo}}},\ }\bibfield  {title} {\bibinfo {title} {Currents induced by
  electron motion},\ }\href {https://doi.org/10.1109/JRPROC.1939.228757}
  {\bibfield  {journal} {\bibinfo  {journal} {Proceedings of the IRE}\ }\textbf
  {\bibinfo {volume} {27}},\ \bibinfo {pages} {584} (\bibinfo {year}
  {1939})}\BibitemShut {NoStop}%
\bibitem [{\citenamefont {Reiser}(2008)}]{reiser1994theory}%
  \BibitemOpen
  \bibfield  {author} {\bibinfo {author} {\bibfnamefont {M.}~\bibnamefont
  {Reiser}},\ }\href {https://doi.org/10.1002/9783527622047} {\emph {\bibinfo
  {title} {Theory and Design of Charged Particle Beams}}}\ (\bibinfo
  {publisher} {Wiley-VCH},\ \bibinfo {year} {2008})\BibitemShut {NoStop}%
\bibitem [{\citenamefont {Buon}(1992)}]{buon1992beam}%
  \BibitemOpen
  \bibfield  {author} {\bibinfo {author} {\bibfnamefont {J.}~\bibnamefont
  {Buon}},\ }\href
  {https://inis.iaea.org/search/search.aspx?orig_q=RN:24057180} {\emph
  {\bibinfo {title} {Beam phase space and emittance (LAL-RT--92-03)}}},\
  \bibinfo {type} {Tech. Rep.}\ (\bibinfo  {institution} {Paris-11 Univ.},\
  \bibinfo {year} {1992})\BibitemShut {NoStop}%
\bibitem [{\citenamefont {Zhang}\ \emph {et~al.}(2017)\citenamefont {Zhang},
  \citenamefont {Valfells}, \citenamefont {Ang}, \citenamefont {Luginsland},\
  and\ \citenamefont {Lau}}]{doi:10.1063/1.4978231}%
  \BibitemOpen
  \bibfield  {author} {\bibinfo {author} {\bibfnamefont {P.}~\bibnamefont
  {Zhang}}, \bibinfo {author} {\bibfnamefont {A.}~\bibnamefont {Valfells}},
  \bibinfo {author} {\bibfnamefont {L.~K.}\ \bibnamefont {Ang}}, \bibinfo
  {author} {\bibfnamefont {J.~W.}\ \bibnamefont {Luginsland}},\ and\ \bibinfo
  {author} {\bibfnamefont {Y.~Y.}\ \bibnamefont {Lau}},\ }\bibfield  {title}
  {\bibinfo {title} {100 years of the physics of diodes},\ }\href
  {https://doi.org/10.1063/1.4978231} {\bibfield  {journal} {\bibinfo
  {journal} {Applied Physics Reviews}\ }\textbf {\bibinfo {volume} {4}},\
  \bibinfo {pages} {011304} (\bibinfo {year} {2017})},\ \Eprint
  {https://arxiv.org/abs/https://doi.org/10.1063/1.4978231}
  {https://doi.org/10.1063/1.4978231} \BibitemShut {NoStop}%
\end{thebibliography}%
\end{document}